\documentclass[11pt,a4paper]{article}

\usepackage{amsmath}
\usepackage{amssymb}
\usepackage{appendix}
\usepackage{multirow}
\usepackage{geometry}
\usepackage{float}
\usepackage{cite}

\usepackage{graphicx,axodraw}

\allowdisplaybreaks

\renewcommand{\Re}{\mathrm{Re}}
\renewcommand{\Im}{\mathrm{Im}}

\begin{document}
\thispagestyle{empty}

\begin{flushright}
LPT-ORSAY 09-107\\
CPT-P082-2009\\
CAS-IHEP/09-11
\end{flushright}

\vspace{\baselineskip}

\begin{center}
\vspace{3\baselineskip}
\textbf{\Large Extracting CP violation and strong phase in $D$ decays\\[0.5em] by using quantum correlations in\\[0.5em]
$\psi(3770)\to D^0\bar{D}^0\to (V_1V_2)(V_3V_4)$ and $\psi(3770)\to
D^0\bar{D}^0 \to (V_1V_2)(K\pi)$
}\\

 \vspace{3\baselineskip}
 {\sc J\'er\^ome Charles$^b$, S\'ebastien Descotes-Genon$^c$, Xian-Wei Kang$^{a,d}$, Hai-Bo
 Li$^a$ and Gong-Ru Lu$^d$
}\\

\vspace{1.4cm}
 {\sl $^a$ Institute of High Energy Physics,

P.O. Box 918, Beijing 100049, China}
 \vspace{\baselineskip}

  {\sl $^b$ Centre de Physique
Th\'eorique~\footnote{Laboratoire affili\'e \`a la FRUMAM} ,

CNRS \& Univ. Aix-Marseille 1 \& 2 and Sud Toulon-Var (UMR
6207),

Luminy Case 907, 13288 Marseille Cedex 9, France}
\vspace{\baselineskip}

{\sl $^c$ Laboratoire de Physique Th\'eorique,

CNRS \&Univ. Paris-Sud 11 (UMR 8627), 91405 Orsay Cedex, France}
\vspace{\baselineskip}

{\sl $^d$ Department of Physics, Henan Normal University,

Xinxiang 453007, China}

\vspace{\baselineskip} \vspace*{0.5cm}

\textbf{Abstract}\\

\end{center}
The charm quark offers interesting opportunities to cross-check
the mechanism of CP violation precisely tested in the strange and
beauty sectors. In this paper, we exploit the angular and quantum
correlations in the $D\bar{D}$ pairs produced through the decay of
the $\psi(3770)$ resonance in a charm factory to investigate
CP-violation in two different ways. We build CP-violating
observables in $\psi(3770)\rightarrow D\bar{D}\rightarrow
(V_1V_2)(V_3 V_4)$ to isolate specific New Physics effects in the
charm sector. We also consider the case of $\psi(3770)\rightarrow
D\bar{D}\rightarrow (V_1V_2)(K\pi)$ decays, which provide a new
way to measure the strong phase difference $\delta$ between
Cabibbo-favoured and doubly-Cabibbo suppressed $D$ decays required
in the determination of the CKM angle $\gamma$. Neglecting the
systematics, we give a first rough estimate of the sensitivities
of these measurements  at BES-III with an integrated luminosity of
20 fb$^{-1}$ at $\psi(3770)$ peak and at a future Super
$\tau$-charm factory with a luminosity of $10^{35}$
cm$^{-2}$s$^{-1}$.
\vspace{1\baselineskip}

\parbox{0.9\textwidth}{

}


\clearpage


\newpage
\setcounter{page}{1}

\section{Introduction}
\label{introduction}
Outstanding progress has been made over the last decade thanks to
the data gathered at $B$-factories, confirming that the
Cabibbo-Kobayashi-Maskawa (CKM) mechanism embedded in the Standard
Model (SM) is the main source of CP violation in the quark sector.
The impressive agreement between results from the $s$-quark and
the $b$-quark
sectors~\cite{Charles:2004jd,Bona:2006ah}
calls for further checks in less tested areas. The recent
discussions concerning the leptonic decays of $D$ and $D_s$
mesons, about a possible disagreement between lattice results and
experimental data~\cite{Dobrescu:2008er,Alexander:2009ux,:2007ws},
suggest that the charm sector has not been explored as extensively as
other quarks~\cite{charmNP}. Another illustration of this situation stems from
$D$-meson mixing, which has only very recently provided
interesting tests of the SM and its
extensions~\cite{Aubert:2007wf,Staric:2007dt,Golowich:2007ka,liyang:2007,liyang:2006}.

Indeed, the D-meson sector is a remarkable place to improve our
knowledge on CP violation in and beyond SM, for at least two
different reasons. First, the SM predictions for CP violation in
the charm sector are very small, due to the hierarchical structure
of the CKM matrix and the difference of masses between the fermion
generations. Any significant amount of CP violation would provide
clear signals of New Physics, and a contrario, the absence of
observation of CP violation already sets bounds on models beyond
the SM~\cite{charmNP}. Secondly, $D$ decays
play a prominent role in determining $\gamma$, the least well known of the
three angles from the $B$-meson unitarity triangle. A better
understanding of the strong dynamics of related $D$ decays would
help in reducing the current uncertainty on this
angle~\cite{Gronau:2001nr,Atwood:2003mj,Asner:2005wf}.

On the experimental side, the final results from CLEO-c and the
start of BES-III provide interesting opportunities. These charm factories
are known to offer the
possibility to exploit the quantum entanglement of $D\bar{D}$
pairs, as explained in several
references~\cite{Bigi:1986dp,Bigi:1989ah,Xing:1996pn,Gronau:2001nr,
Asner:2005wf,quantumDDbar}. In addition, it is also interesting to note
that $D\to VV$ (vector-vector)
modes exhibit rather large branching ratios, of similar size with
respect to
the pseudoscalar ($PP$) or vector-pseudoscalar ($VP$) modes, and provide
further angular observables to study the above
issues. In this paper, we investigate this question which had not been
detailed so far.

In Section~\ref{sec:2}, we discuss production of coherent
$D^0$-$\bar{D}^0$ pairs from $\psi(3770)$ decay, and in particular
the angular distribution when at least one of the $D$ meson decays
into a pair of vector mesons. In Section~\ref{sec:3}, we apply
these results to two different situations: the determination of
CP-violating observables exploiting angular and quantum
correlations in cases where both $D$ decay into vector pairs, and
the extraction of $D\to K\pi$ hadronic parameters in relation with
the measurement of the CKM angle $\gamma$. In Section~\ref{sec:4},
we briefly discuss the application of these results for BES-III
and Super $\tau$-charm factory, before concluding.

\section{Correlated $D$ decays}
\label{sec:2}
\subsection{Basics}~\label{sec:basics}\label{sec:BRs}
\label{sec:2:subsec:1}
 We want to describe the decay chain for
$\psi(3770)$ as
\begin{equation}
\psi(3770) \to D^0\bar{D}^0\to (M_1M_2) (M_3M_4).
\end{equation}
where $M_1 M_2$ and $M_3M_4$ are mesons from two-body decays of
$D^0$ and $\bar{D}^0$, respectively (Hereafter, $\psi$ denotes
$\psi(3770)$). Since we do not tag the $D$ mesons, and just
observe their decay products, we can use two descriptions of $D$
mesons -- either the flavour states $D^0$ and $\bar{D}^0$, or the
CP eigenstates (neglecting, for the sake of simplicity, CP-violation in
$D$ mixing):
\begin{equation}
|D_1\rangle= \frac{|D^0\rangle + |\bar{D}^0\rangle}{\sqrt{2}},
\qquad |D_2\rangle= \frac{|D^0\rangle -
|\bar{D}^0\rangle}{\sqrt{2}}
\end{equation}
with respective CP-parity: $\eta_{CP}(D_1)=-1$ and
$\eta_{CP}(D_2)=1$ (we take the
convention $CP|D^0\rangle=-|\bar{D}^0\rangle $).

Due to the spin of $\psi$, the $D$-pair
is emitted with an orbital momentum $L=1$ corresponding to
an antisymmetric coherent state:
\begin{equation}
|(D\bar{D})_{L=1}\rangle=\frac{-|D_1\rangle|D_2\rangle+|D_2\rangle|D_1\rangle}{\sqrt{2}}.
\end{equation}

One can in principle consider different situations as below, where $V$
stands for a vector and $P$ for a pseudoscalar meson:
\begin{itemize}
\item $(PP)+(PP),(PP)+(VP),(VP)+(VP)$: the only available observable is the
branching ratio, since the partial waves and helicities are all
fixed by angular momentum conservation.
\item $(PP)+(VV),(VP)+(VV)$: $(VV)$ can have three helicity states, and thus there are new angular observables. This can be exploited for $(PP)=K\pi$ in connection with the measurement of the CKM angle $\gamma$.
\item $(VV)+(VV)$: this will be studied with an interest in new observables for CP-violation.
\end{itemize}

The relevant modes for our studies can be extracted from
Ref.~\cite{Amsler:2008zzb} for the branching ratios and
Ref.~\cite{Asner:2008nq} for the projected efficiency at BES-III.

\begin{table}[htp]
\begin{center}
\begin{tabular}{cccc}\hline\hline
$PP$    &   $\eta_{CP}(P)\eta_{CP}(P)$ &    Br (\%) & Eff. ($\epsilon$) \\
\hline
$K^+K^-$    &   +1  & 0.39  & 0.50  \\
$\pi^+\pi^-$    &   +1  & 0.14  & 0.60  \\
$K_S K_S$   &   +1  & 0.038 & 0.30  \\
$\pi^0\pi^0$    &   +1  & 0.08  & 0.24  \\
$K_S\pi^0$  &   -1  & 1.22  & 0.33  \\
$K_S \eta$  &   -1  & 0.40  & 0.26 \\
$K_S a_0(980)\to K_S(\eta\pi^0)$ & +1 & 0.67 & 0.18 \\
$K_S a_0(980) \to K_S(K^+K^-)$ &  +1 &  0.31 & 0.10\\\hline\hline
\end{tabular}
\caption{Branching ratios for $D$ decays into CP-eigenstates
composed of two pseudoscalar mesons. In each case, the product of
intrinsic $CP$ parities and the estimated reconstruction
efficiency at BES-III are indicated. Note that the efficiency is
for both $D^0$ decaying into a $PP$ final state; for single $D$
decay, the efficiency is $\sqrt{\epsilon}$. } \label{tab:1}
\end{center}
\end{table}

\begin{table}[H]
\begin{center}
\begin{tabular}{cccc}\hline\hline
$PV$    &   $\eta_{CP}(P)\eta_{CP}(V)$  &   Br (\%) & Eff. ($\epsilon$)\\
\hline
$\rho^0\pi^0$   &   -1  & 0.37  & 0.29  \\
$\phi \pi^0 \to (K^+K^-)\pi^0$& -1 & 0.06 & 0.10 \\
$K_S \rho^0$    &   +1  & 0.77  & 0.27  \\
$K_S \phi\to K_S (K^+K^-)$  &   +1  & 0.22  & 0.08 \\
$K_S \omega\to K_S (\pi^+ \pi^- \pi^0)$     &   +1  & 0.98  & 0.20 \\
$\bar K^{*0} \eta    \to (K_S \pi^0) (\pi^+ \pi^- \pi^0)$ &  +1  & 0.03  & 0.17\\
$\bar K^{*0} \eta    \to (K_S \pi^0) (\gamma\gamma)$ &   +1  & 0.06  & 0.17\\
$\bar{K}^{*0}\pi^0\to(K_S\pi^0)\pi^0$ & +1 &  0.67 &
0.15\\\hline\hline
\end{tabular}
\caption{Branching ratios for $D$ decays into CP-eigenstates
composed of one pseudoscalar and one vector mesons. In each case,
the product of intrinsic $CP$ parities and the estimated
reconstruction efficiency at BES-III are indicated. Note that the
efficiency is for both $D^0$ decaying into a $PV$ final state; for
single $D$ decay, the efficiency is
$\sqrt{\epsilon}$.}\label{tab:2}
\end{center}
\end{table}

\begin{table}[H]
\begin{center}
\begin{tabular}{cccc}\hline\hline
$VV$        &$\eta_{CP}(V)\eta_{CP}(V)$  &   Br (\%) & Eff. ($\epsilon$)\\
\hline
$\rho^0 \rho^0$ &   1   &0.18   & 0.24 \\
$\bar{K}^{*0} \rho^0 \to (K_S \pi^0) (\pi^+\pi^-)$ & 1  &0.27   &0.12   \\
$\rho^0 \phi \to (\pi^+\pi^-)(K^+K^-)$   &   1   &0.14   & 0.07\\
$\bar{K}^{*0} \omega \to (K_S \pi^0) (\pi^+\pi^-\pi^0)$ &   1   &0.33   &0.09\\
$\rho^+ \rho^-$ &   1   & [0.6]    &0.18\\
$\rho^0 \omega\to (\pi^+ \pi^-) (\pi^+\pi^-\pi^0)$  &   1   & [$\simeq$ 0]  &0.18\\
$K^{*+} K^{*-}\to (K_S \pi^+)(K_S \pi^-)$   &   1   & [0.08]    &0.07\\
$K^{*0} \bar{K}^{*0}\to (K_S \pi^0)(K_S \pi^0)$ &   1   &0.003  &0.09\\
 \hline\hline
\end{tabular}
\caption{Branching ratios for $D$ decays into CP-eigenstates
composed of two vector mesons. In each case, the product of
intrinsic $CP$ parities and the estimated reconstruction
efficiency at BES-III are indicated. The rates in brackets are not
measured yet, but were predicted in ref.~\cite{Uppal:1992se}.
 Note that the efficiency is for both $D^0$ decaying into a $VV$ final
state; for single $D$ decay, the efficiency is
$\sqrt{\epsilon}$.}\label{tab:3}
\end{center}
\end{table}

 Some of the $VV$ modes have not been measured yet, but
some estimates combining naive factorisation and models for
final-state interactions are
available~\cite{Uppal:1992se,Kamal:1991,Bedaque:1994,Hinchliffe:1996}.
It is interesting to notice that the $\rho^+\rho^-$ decay mode has
not been measured yet, even though one would expect it to be
larger than $\rho^0\rho^0$ (the latter being a colour suppressed
mode).

As it will become clear below, for the modes of definite CP-parity, one
can sum over all the possible subsequent decay
channels (enhancing therefore the branching ratios).
For modes without such definite CP-parity, one needs to pick up a subsequent
decay channel providing a definite CP parity. For instance, in the case of the modes
where $K^{*0}$ is identified through the channel $K^{*0} \to K_S \pi^0$,
we have included a factor 1/6 in the branching ratio
due to Clebsch-Gordan coefficients ($1/\sqrt{3}$ from $K^{*0}$ to $K^0\pi^0$,
then $1/\sqrt{2}$ from $K^0$ to $K_S$).

Only the product of intrinsic parity is given, and one has to include the
partial wave of the outgoing state. If $PP$ as well as $VV$ in $S$ and
$D$ waves are not affected, one has to include an additional $(-1)$ for $P$-wave states:
\begin{equation}
\eta_{CP}(PV)=-\eta_{CP}(P)\eta_{CP}(V),\qquad
\eta(VV,\ell=1)=-\eta_{CP}(V)\eta_{CP}(V).
\end{equation}

\subsection{Differential decay width}
\label{sec:2:subsec:2}

 An adequate formalism to treat the question of
decay chains is the framework of helicity amplitudes, described for
instance in Refs.~\cite{Jackson,Jacob}. The decay chain is described
by the product of amplitudes corresponding to each reaction. For a
reaction $A\to BC$, we define polar angles $(\theta_A,\phi_A)$
describing the momentum of particle $B$ in the rest frame of $A$ in
a basis where the $z$-axis defined by the momentum of $B$ in the
rest frame of its mother particle. The decay amplitude depends on
$(\theta_A,\phi_A)$ and is denoted $A^{A\to
BC}_{\lambda_B\lambda_C}$ where $\lambda_B,\lambda_C$ are the
helicities of the daughter mesons.

Let us start by describing the chain (we will study the other ``path'' later):
\begin{eqnarray}
&& \psi \to D_1 D_2,\qquad
 D_1 \to V_1 V_2, \quad D_2 \to V_3 V_4,\\
&&V_1 \to M_1 M'_1, V_2 \to M_2 M'_2,
  V_3 \to M_3 M'_3, V_4 \to M_4 M'_4.
\end{eqnarray}
The helicity formalism yields an amplitude of the form
\begin{eqnarray}
M^m_{12} &=& \sum_{\lambda_V} A^{\psi\to D_1 D_2}_{00} A^{V_1\to M_1M_1'}_{00}
         A^{V_3\to M_3M_3'}_{00}
A^{V_2\to M_2M_2'}_{00}
         A^{V_4\to M_4M_4'}_{00}
A^{D_1\to V_1V_2}_{\lambda_{V_1}\lambda_{V_2}}
 A^{D_2\to V_3V_4}_{\lambda_{V_3}\lambda_{V_4}}\\
&=& \sqrt{\frac{3}{4\pi}} \frac{9}{(4\pi)^3}
 \sum_{\lambda_V} D^{1*}_{m,0}(\phi_\psi,\theta_\psi,0) H^\psi_{D_1D_2}
\\ \nonumber
&&\times
     D^{0*}_{0,\lambda_{V_1}-\lambda_{V_2}}(\phi_{D_1},\theta_{D_1},0) H^{D_1}_{V_1V_2}
      D^{1*}_{\lambda_{V_1},0}(\phi_{V_1},\theta_{V_1},0) H^{V_1}_{M_1M'_1}
     D^{1*}_{-\lambda_{V_2},0}(\phi_{V_2},\theta_{V_2},0) H^{V_2}_{M_2M'_2}\\ \nonumber
&&\times
     D^{0*}_{0,\lambda_{V_3}-\lambda_{V_4}}(\phi_{D_2},\theta_{D_2},0) H^{D_2}_{V_3V_4}
     D^{1*}_{\lambda_{V_3},0}(\phi_{V_3},\theta_{V_3},0) H^{V_3}_{M_3M'_3}
     D^{1*}_{-\lambda_{V_4},0}(\phi_{V_4},\theta_{V_4},0)
     H^{V_4}_{M_4M'_4},
\end{eqnarray}
where $m$ is the projection of the spin of the $\psi$ along an arbitrary axis,
and $\lambda_V$ denotes collectively the helicities of the 4 vector
mesons. The vector mesons are emitted from a spinless $D$-meson, so that:
$\lambda=\lambda_{V_1}=\lambda_{V_2}$
and $\kappa=\lambda_{V_3}=\lambda_{V_4}$.
We used the rotation matrix
$D^j_{m'm}(\alpha,\beta,\gamma)=e^{-im'\alpha}d^j_{m'm}(\beta)e^{-im\gamma}$
with the Wigner $d$-matrix:
\begin{equation}
d^1_{10}(\theta)=-\frac{1}{\sqrt{2}}\sin\theta,\quad
d^1_{00}(\theta)=\cos\theta,\quad
d^1_{-10}(\theta)=\frac{1}{\sqrt{2}}\sin\theta,\quad
d^0_{00}(\theta)=1.
\end{equation}

The probability amplitude becomes
\begin{eqnarray}
&&M^m_{12}
= \sqrt{\frac{3}{4\pi}} \frac{9}{(4\pi)^3} e^{im\phi_\psi}
d^1_{m0}(\theta_\psi) H^{\psi V_1 V_2 V_3 V_4}\\ \nonumber
&&\times
  \sum_{\lambda} e^{i\lambda \Phi_{12}} (-1)^\lambda d^1_{\lambda 0}(\theta_{V_1})
  d^1_{\lambda 0}(\theta_{V_2}) H^{D_1}_{\lambda}
  \sum_{\kappa} e^{i\kappa \Phi_{34}} (-1)^{\kappa} d^1_{\kappa 0}(\theta_{V_3})
  d^1_{\kappa 0}(\theta_{V_2}) H^{D_2}_{\kappa}\\ \nonumber
&&= \sqrt{\frac{3}{4\pi}} \frac{9}{(4\pi)^3} e^{im\phi_\psi}
d^1_{m0}(\theta_\psi) H^{\psi V_1 V_2 V_3 V_4}\\ \nonumber
&&\times
 \Bigg[\cos\theta_{V_1}\cos\theta_{V_2} A^{D_1\to V_1V_2}_0
  -\frac{1}{\sqrt{2}}\sin\theta_{V_1}\sin\theta_{V_2}\cos\Phi_{12} A^{D_1\to
    V_1V_2}_{||}
  -\frac{i}{\sqrt{2}}\sin\theta_{V_1}\sin\theta_{V_2}\sin\Phi_{12} A^{D_1\to
    V_1V_2}_{\perp}
  \Bigg]\\ \nonumber
&&\times
 \Bigg[\cos\theta_{V_3}\cos\theta_{V_4} A^{D_2\to V_3V_4}_0
  -\frac{1}{\sqrt{2}}\sin\theta_{V_3}\sin\theta_{V_4}\cos\Phi_{34} A^{D_2\to
    V_3V_4}_{||}
  -\frac{i}{\sqrt{2}}\sin\theta_{V_3}\sin\theta_{V_4}\sin\Phi_{34} A^{D_2\to
    V_3V_4}_{\perp}
  \Bigg],
\end{eqnarray}
where we defined $\Phi_{12}=\phi_{V_1}-\phi_{V_2}$ and
$\Phi_{34}=\phi_{V_3}-\phi_{V_4}$ (i.e. the angle between the two
relevant vector mesons) and the combination of amplitudes
\begin{eqnarray}
H^{\psi V_1 V_2 V_3 V_4} &=&H^\psi_{D_1D_2}
     H^{V_1}_{M_1M_1'}H^{V_2}_{M_2M_2'}H^{V_3}_{M_3M_3'}H^{V_4}_{M_4M_4'},\\
H^{D_1\to V_1V_2}_{\lambda}&=&H^{D_1\to V_1V_2}_{\lambda\lambda},
\qquad H^{D_2\to V_3V_4}_{\kappa}=H^{D_2\to V_3V_4}_{\kappa\kappa}.
\end{eqnarray}
We introduce the transversity amplitudes
\begin{equation}
A_{||}=\frac{1}{\sqrt{2}}(H_{+1}+H_{-1}),\qquad
A_{0}=H_0,\qquad
A_\perp=\frac{1}{\sqrt{2}}(H_{+1}-H_{-1}).
\end{equation}

$M^m_{12}$ is actually only one of the two ``paths'' that can be chosen. The total
amplitude for a given projection $m$ of the spin of $\psi$ along an arbitrary
$z$-axis is: $M^m=(-M^m_{12}+M^m_{21})/\sqrt{2}$.
The differential decay width is obtained by averaging the squared modulus
of
the amplitude over the three possible values of $m=+1,0,-1$. The three
squared Wigner functions $d^1_{m0}(\theta_\psi)$ add up to 1, so that the
differential width is
\begin{eqnarray}\label{Gamma4V}
&&d\Gamma_{4V}=  \frac{81}{32\pi^2}
 d(\cos\theta_{V_1}) d(\cos\theta_{V_2}) d\Phi_{12}
 d(\cos\theta_{V_3}) d(\cos\theta_{V_4}) d\Phi_{34}
\times|A^{\psi V_1 V_2 V_3
  V_4}|^2\\ \nonumber
&&\times \Bigg|\Bigg[\cos\theta_{V_1}\cos\theta_{V_2} A^{D^0\to
V_1V_2}_0
  -\frac{1}{\sqrt{2}}\sin\theta_{V_1}\sin\theta_{V_2}\cos\Phi_{12} A^{D^0\to
    V_1V_2}_{||}
  -\frac{i}{\sqrt{2}}\sin\theta_{V_1}\sin\theta_{V_2}\sin\Phi_{12} A^{D^0\to
    V_1V_2}_{\perp}
  \Bigg]\\ \nonumber
&&\times
 \Bigg[\cos\theta_{V_3}\cos\theta_{V_4} A^{\bar{D}^0\to V_3V_4}_0
  -\frac{1}{\sqrt{2}}\sin\theta_{V_3}\sin\theta_{V_4}\cos\Phi_{34} A^{\bar{D}^0\to
    V_3V_4}_{||}
  -\frac{i}{\sqrt{2}}\sin\theta_{V_3}\sin\theta_{V_4}\sin\Phi_{34} A^{\bar{D}^0\to
    V_3V_4}_{\perp}
  \Bigg]\\ \nonumber
&&
 -
 \Bigg[\cos\theta_{V_1}\cos\theta_{V_2} A^{\bar{D}^0\to V_1V_2}_0
  -\frac{1}{\sqrt{2}}\sin\theta_{V_1}\sin\theta_{V_2}\cos\Phi_{12} A^{\bar{D}^0\to
    V_1V_2}_{||}
  -\frac{i}{\sqrt{2}}\sin\theta_{V_1}\sin\theta_{V_2}\sin\Phi_{12} A^{\bar{D}^0\to
    V_1V_2}_{\perp}
  \Bigg]\\ \nonumber
&&\times
 \Bigg[\cos\theta_{V_3}\cos\theta_{V_4} A^{D^0\to V_3V_4}_0
  -\frac{1}{\sqrt{2}}\sin\theta_{V_3}\sin\theta_{V_4}\cos\Phi_{34} A^{D^0\to
    V_3V_4}_{||}
  -\frac{i}{\sqrt{2}}\sin\theta_{V_3}\sin\theta_{V_4}\sin\Phi_{34} A^{D^0\to
    V_3V_4}_{\perp}
  \Bigg]\Bigg|^2.
\end{eqnarray}

We have an integration over $[0,\pi]$ for $\theta$'s and $[0,2\pi]$
for $\Phi_{12}$ and $\Phi_{34}$. The amplitudes $A$ are normalized
so that: $\Gamma(X\to YZ)=|A(X\to YZ)|^2$.

The above formalism can be adapted easily to describe the
situation where one $D$ meson decays into $PP^\prime$ rather than
$VV^\prime$. Indeed, it amounts to considering only the
longitudinal decay amplitude for $D\to V_3 V_4$ and to remove the
angular phase space related to the decay products of $V_3$ and $V_4$.

\section{Observables from correlated $D$ decays}
\label{sec:3}

\subsection{Observables from $\psi\to 2D\to 4V$  for CP violation}
\label{sec:3:subsec:2}

 If we take the decay
chain~\cite{Bigi:1986dp,Bigi:1989ah}
\begin{equation}
e^+e^- \to \psi \to D^0 \bar{D}^0 \to f_a f_b
\end{equation}
with $f_a$ and $f_b$ CP eigenstates of same CP-parity, we have
\begin{equation}
CP|\psi\rangle = |\psi\rangle, \qquad CP|f_af_b\rangle =
\eta_a\eta_b(-1)^\ell|f_a f_b\rangle = -|f_a f_b\rangle
\end{equation}
since $f_a$ and $f_b$ are in a $P$ wave. Therefore, the decay of $\psi$
into
states of identical $CP$ parity is, by itself, a CP-violating
observable~\cite{Bigi:1986dp,Bigi:1989ah}.

One obtains, neglecting CP-violation in $D\bar{D}$ mixing, the
following result for the combined branching ratio, which can be
recovered from~\cite{Xing:1996pn}
\begin{eqnarray} \label{eq:brcp}
 \mathcal{BR}((D^0\bar{D}^0)_{C=-1} \to f_a f_b)= 2\mathcal{BR}(D^0\to
f_a)\mathcal{BR}(D^0\to
f_b) (\left|\rho_a-\rho_b\right|^2+r_D|1-\rho_a\rho_b|^2 ),
\end{eqnarray}
with the ratio of CP-conjugate amplitudes and the combination of $D$-mixing parameters
\begin{equation}\label{rhof}
\rho_f=\frac{A(\bar{D}^0\to f)}{A(D^0\to f)}, \qquad
r_D=(x^2+y^2)/2<10^{-4}.
\end{equation}
where $x=\Delta m/\Gamma$ and $y=\Delta \Gamma/(2\Gamma)$ are the difference of masses
and widths of the mass eigenstates in the $D\bar{D}$ system, normalised by their average width~\cite{charmNP}.

If we assume that CP is conserved in decay, we have
$\rho_f=\eta_f$, and thus $\mathcal{BR}=0$ for $a,b$ with same CP-parity.
Therefore, we have indeed that  the observation of
$(D^0\bar{D}^0)_{C=-1} \to f_a f_b$ with $a,b$ of same CP-parity
is an indication of $CP$-violation. Let us notice that $a$ and $b$
must be different eigenstates (either different mesons, or for
$VV$, different partial waves), and that this branching ratio is
sensitive to different aspects of CP-violation compared to
uncorrelated decays of $D\to f_a$ and $D\to f_b$, since the latter
would be sensitive to $1-|\rho_a|^2$ or  $1-|\rho_b|^2$. We can
thus construct observables for CP violation in $VV$ decays by
considering states with the same $CP$ parity, which depends on the
relative angular momentum between the two mesons. The favourite
channels among the measured ones are $K^+K^-$, $\pi^+\pi^-$,
$K_S\pi^0$,$ \rho^0\pi^0$, $K_S \rho^0$, $\bar{K}^{*0} \rho^0 \to
(K_S \pi^0) (\pi^+\pi^-)$ and $\rho^0 \phi$.

The transversity amplitudes $A$ for $D_{1,2}\rightarrow VV^\prime$
have simple transformation laws under $CP$:
\begin{eqnarray}
A_{0}^{D\to VV'}&\to& + \eta_{CP}(V)\eta_{CP}(V') \eta_{CP}(D) A_{0}^{D\to
\bar{V}\bar{V}'},\\
A_{||}^{D\to VV'}&\to& + \eta_{CP}(V)\eta_{CP}(V') \eta_{CP}(D)
A_{||}^{D\to \bar{V}\bar{V}'},\\
A_{\perp}^{D\to VV'}&\to& -\eta_{CP}(V)\eta_{CP}(V') \eta_{CP}(D)
A_{\perp}^{D\to \bar{V}\bar{V}'}.
\end{eqnarray}

Following eq.~(\ref{eq:brcp}), $CP$ conservation at the level of the
amplitude would require that only two combinations of transversity
amplitudes are allowed: $(0,\perp)$ or $(||,\perp)$.
In terms of partial waves, $0$ and $||$ are combinations of $S$ and $D$
waves, whereas $\perp$ is $P$ wave, which means that $CP$ conservation at
the
level of the amplitude would impose the vector mesons to be
emitted in $(S,D)$ waves on one side and $P$ wave on the other.

Therefore, the following combinations of transversity amplitudes in the
partial differential decay rate can be in principle $CP$ violating
observables:
\begin{equation}
(0,0), \qquad (0,||), \qquad (||,0), \qquad (||,||), \qquad
(\perp,\perp).
\end{equation}

Let us notice that in the case of identical meson pairs in the final
state, there is only one CP-violating configuration that is
available: $(0,||)$, due to the Bose-Einstein statistics.
It seems more interesting to consider two different meson pairs, both
with longitudinal polarization, to get a larger BR. From the above
table, the most interesting modes are $\bar{K}^{*0} \rho^0 \to (K_S
\pi^0) (\pi^+\pi^-)$ and $\rho^0 \rho^0$. It is straightforward
to construct the corresponding CP-violating observable:
\begin{eqnarray}\label{observable1}
&&\int  d\Gamma_{4V}
 \frac{1}{128} (5 \cos^2 \theta_{V_1}-1) (5 \cos^2 \theta_{V_2}-1)  (5
\cos^2 \theta_{V_3}-1)
 (5 \cos^2 \theta_{V_4}-1) \\ \nonumber
&&  \qquad \qquad=
 |A^{\psi V_1 V_2 V_3 V_4}|^2 |A^{D^0\to V_1V_2}_0|^2 |A^{D^0\to
V_3V_4}_0|^2
     \times |\rho^0_{V_1,V_2}-\rho^0_{V_3,V_4}|^2.
\end{eqnarray}

Similar weights can be obtained for the other CP-violating combinations,
exploiting orthogonality relationships for Legendre and Chebyshev
polynomials to select specific angular  dependences as in the previous
case. For instance, we have:
\begin{eqnarray}
&&\int  d\Gamma_{4V}\label{observable2}
 \frac{1}{32} (5\cos^2 \theta_{V_1}-3) (5 \cos^2 \theta_{V_2}-3)  (5
\cos^2 \theta_{V_3}-3)
 (5 \cos^2 \theta_{V_4}-3)\\
\nonumber&&\qquad\qquad\cdot(4\cos^2\Phi_{12}-1)(4\cos^2\Phi_{34}-1) \\
\nonumber
&& =
 |A^{\psi V_1 V_2 V_3 V_4}|^2 |A^{D^0\to V_1V_2}_{||}|^2 |A^{D^0\to
V_3V_4}_{||}|^2
     \times |\rho^{||}_{V_1,V_2}-\rho^{||}_{V_3,V_4}|^2
\end{eqnarray}
and
\begin{eqnarray}\label{observable3}
&&\int  d\Gamma_{4V}
 \frac{1}{32} (5\cos^2 \theta_{V_1}-3) (5 \cos^2 \theta_{V_2}-3)  (5
\cos^2 \theta_{V_3}-3)
 (5 \cos^2 \theta_{V_4}-3)\\ \nonumber
 &&\qquad\qquad\cdot(4\cos^2\Phi_{12}-3)(4\cos^2\Phi_{34}-3)\\
 \nonumber
&& = |A^{\psi V_1 V_2 V_3 V_4}|^2 |A^{D^0\to V_1V_2}_{\perp}|^2
|A^{D^0\to V_3V_4}_{\perp}|^2\times
|\rho^{\perp}_{V_1,V_2}-\rho^{\perp}_{V_3,V_4}|^2.
\end{eqnarray}

In the case of the same $VV$ final state for both $D$ decays, one can
obtain the appropriate observable
corresponding to $(0,||)$ using for instance
\begin{eqnarray}
&&\int  d\Gamma_{4V}\label{observable4}
 [(5\cos^2 \theta_{V_1}-1) (5 \cos^2 \theta_{V_2}-1)  (5
\cos^2\theta_{V_3}-3)
 (5 \cos^2 \theta_{V_4}-3)(4\cos^2\Phi_{12}-1) \\ \nonumber
&& \qquad \cdot[(5\cos^2 \theta_{V_1}-3) (5 \cos^2 \theta_{V_2}-3)
(5 \cos^2 \theta_{V_3}-1) (5 \cos^2
\theta_{V_4}-1)(4\cos^2\Phi_{34}-1)]\\ \nonumber &&  \qquad \qquad=
 |A^{\psi V_1 V_2 V_3 V_4}|^2 |A^{D^0\to V_1V_2}_{0}|^2
 |A^{D^0\to V_3V_4}_{||}|^2
     \times |\rho^{0}_{V_1,V_2}-\rho^{||}_{V_3,V_4}|^2.
\end{eqnarray}

\subsection{$\psi\to 2D\to (VV)(K\pi)$ for the extraction of $\gamma$}
\label{sec:3:subsec:1}

The measurement of $\gamma$ from the Atwood-Dunietz-Soni (ADS)
method~\cite{Atwood:1996ci} requires the determination of the
hadronic parameters $r$ and $\delta$. At BES-III, we can also take
advantage of the coherence of the $D^0$ mesons produced at the
$\psi(3770)$ peak to extract the strong phase difference $\delta$
between doubly-Cabibbo-suppressed and Cabibbo-favoured
decay amplitudes that appears in the $\gamma$
measurements~\cite{Gronau:2001nr,Asner:2005wf}. Here we introduce, in the
standard phase convention where $\delta$ vanishes in the SU(3) limit,
\begin{equation}\label{eq:rdeltaforkpi}
r\cdot  e^{i\delta}=\frac{\langle
K^-\pi^+|\bar{D}^0\rangle}{\langle K^-\pi^+|D^0\rangle}.
\end{equation}
 The process of one $D^0$
decaying to $K^-\pi^+$, while the other $D^0$ decaying to a $VV$ $CP$
eigenstate can be described as (for our purposes, it will prove
more convenient to express this decay rate in terms of $D^0$ and
$\bar{D}^0$ amplitudes):
\begin{eqnarray} \label{eq:gamma2V}
d\Gamma_{2V}&=&  \frac{9}{4\pi}
 d(\cos\theta_{V_1}) d(\cos\theta_{V_2}) d\Phi
\times|A^{\psi V_1 V_2}|^2 |A^{D^0\to K\pi}|^2\\ \nonumber
&&\times \Bigg|\cos\theta_{V_1}\cos\theta_{V_2}
  (A^{\bar{D}^0\to V_1V_2}_0-r e^{i\delta}A^{D^0\to V_1V_2}_0)\\ \nonumber
&&\qquad
  -\frac{1}{\sqrt{2}}\sin\theta_{V_1}\sin\theta_{V_2}\cos\Phi
   (A^{\bar{D}^0\to V_1V_2}_{||}-r e^{i\delta}A^{D^0\to V_1V_2}_{||})\\ \nonumber
&&\qquad
  -\frac{i}{\sqrt{2}}\sin\theta_{V_1}\sin\theta_{V_2}\sin\Phi
  (A^{\bar{D}^0\to V_1V_2}_\perp-r e^{i\delta}A^{D^0\to V_1V_2}_\perp)
  \Bigg|^2.
\end{eqnarray}
We can introduce:
\begin{equation}
A_{0,||,\perp}(\bar{D}^0\to V_a V_b)=A_{0,||,\perp}(D^0\to V_a
V_b)\rho^{0,||,\perp}_{V_a,V_b}.
\end{equation}
In the absence of $CP$ violation, which we will assume in this section, we
have:
\begin{equation}
\rho^{0,||}_{V_a,V_b}=-\eta_{CP}(V_a)\eta_{CP}(V_b)=-\rho^{\perp}_{V_a,V_b
} .
\end{equation}
Moreover, we notice that all the decays presented in
Sec.~\ref{sec:BRs} have CP parities such that $\rho^0=-1$, which
yields the further expression of the differential decay width in
Eq.(\ref{eq:gamma2V})
\begin{eqnarray} \label{eq:angdist}
d\Gamma_{2V} &=&  \frac{9}{4\pi}
 d(\cos\theta_{V_1}) d(\cos\theta_{V_2}) d\Phi
\times  |A^{\psi V_1 V_2}|^2 |A^{D^0\to K\pi}|^2 \\ \nonumber
&&\times \Big[\cos^2\theta_{V_1}\cos^2\theta_{V_2}
  |A^{D^0\to V_1V_2}_0|^2(1+2r\cos\delta+r^2)\\ \nonumber
&&\qquad +\frac{1}{2}\sin^2\theta_{V_1}\sin^2\theta_{V_2} \cos^2\Phi
  |A^{D^0\to V_1V_2}_{||}|^2(1+2r\cos\delta+r^2)\\ \nonumber
&&\qquad  -\sqrt{2}\cos\theta_{V_1}\sin\theta_{V_1}\cos\theta_{V_2}\sin\theta_{V_2} \cos\Phi
  \Re[A^{D^0\to V_1V_2}_0(A^{D^0\to V_1V_2}_{||})^*](1+2r\cos\delta+r^2)\\
\nonumber
&&\qquad +\frac{1}{2} \sin^2\theta_{V_1}\sin^2\theta_{V_2} \sin^2\Phi
  |A^{D^0\to V_1V_2}_\perp|^2(1-2r\cos\delta+r^2)\\ \nonumber
&&\qquad  +\sqrt{2}\cos\theta_{V_1}\sin\theta_{V_1}\cos\theta_{V_2}\sin\theta_{V_2} \sin\Phi
  \Big\{\Re[A^{D^0\to V_1V_2}_0(A^{D^0\to
V_1V_2}_\perp)^*](2r\sin\delta)\\
\nonumber
&&\qquad   +\Im[A^{D^0\to V_1V_2}_0(A^{D^0\to
V_1V_2}_\perp)^*](1-r^2)\Big\}\\ \nonumber
&&\qquad  -\sin^2\theta_{V_1}\sin^2\theta_{V_2}\cos\Phi \sin\Phi
  \Big\{\Re[A^{D^0\to V_1V_2}_{||}(A^{D^0\to
V_1V_2}_\perp)^*](2r\sin\delta)\\ \nonumber
&&\qquad   +\Im[A^{D^0\to V_1V_2}_{||}(A^{D^0\to
V_1V_2}_\perp)^*](1-r^2)\Big\}
\Big].
\end{eqnarray}

We see that the differential decay width provides six different angular
observables depending on the following (real) quantities:
\begin{itemize}
\item three products of moduli for $VV$ decays:  $|A^{\psi V_1 V_2}
A^{D^0\to K\pi} A^{D^0\to V_1V_2}_{0,||,\perp}|^2 $
\item two
relative phases between the three amplitudes $A^{D^0\to
V_1V_2}_{0,||,\perp}$
\item two strong parameters describing the
$K\pi$ decay: $r$ and $\delta$
\end{itemize}

 Whereas the full angular integration yields the sum of the three transversity amplitudes:
\begin{eqnarray}\label{Gamma2V:full integration}
 \int d\Gamma_{2V} &=&\frac{9}{4\pi}|A^{\psi
V_1 V_2}|^2 |A^{D^0\to K\pi}|^2 \Big[\frac{8\pi}{9}|A^{D^0\to
V_1V_2}_0|^2(1+2r\cos\delta+r^2)+\\\nonumber
&&\frac{8\pi}{9}|A^{D^0\to V_1V_2}_{||}|^2(1+2r\cos\delta+r^2)
+\frac{8\pi}{9}|A^{D^0\to
V_1V_2}_\perp|^2(1-2r\cos\delta+r^2)\Big],
\end{eqnarray}
one can easily separate the different contributions by choosing suitable
weights for the
angular integration (they can be obtained easily by exploiting
orthogonality relations among Legendre polynomials). In practice the best
way to perform the experimental analysis is usually to do a
maximum likelihood fit on Eq.~(\ref{eq:angdist}).

\begin{table}[!htp]
\begin{equation*}
\begin{array}{ccc}\hline\hline
i & P_i(\theta_{V_1},\theta_{V_2},\Phi) & \int d\Gamma_{2V} P_i/(|A^{\psi V_1 V_2}|^2 |A^{D^0\to K\pi}|^2)\\
\hline 1 & \frac{1}{8} (5 \cos\theta_{V_1}^2-1) (5
\cos\theta_{V_2}^2-1) &
|M_0|^2 \\
2 & \frac{1}{16} (5 \cos\theta_{V_1}^2-3) (5 \cos\theta_{V_2}^2-3)
(4 \cos\Phi^2-1)  &
|M_{||}|^2 \\
3&-\frac{25}{4\sqrt{2}}\cos\theta_{V_1}\cos\theta_{V_2}\sin\theta_{V_1}
\sin\theta_{V_2}\cos\Phi &
\Re\left[M_0M_{||}^*\right] \\
4 & -\frac{1}{16}  (5 \cos\theta_{V_1}^2-3) (5
\cos\theta_{V_2}^2-3) (4 \cos\Phi^2-3) & 
|M_\perp|^2 \\
5 & \frac{25}{4\sqrt{2}} \cos\theta_{V_1}\sin\theta_{V_1}
\cos\theta_{V_2}   \sin\theta_{V_2} \sin\Phi &
-\Re\left[M_0 M_\perp^*\right] \\
6 & \frac{1}{4}  (5 \cos\theta_{V_1}^2-3) (5
\cos\theta_{V_2}^2-3)\cos\Phi\sin\Phi  & 
\Re\left[M_{||} M_\perp^*\right] \\
\hline\hline
\end{array}
\end{equation*}
\caption{Weights used to select contributions from the transversity
amplitudes for $\psi\to (K\pi)(VV)$. $M$ amplitudes are defined by
Eq.~(\ref{eq:Ms}).} \label{kpi:weight}
\end{table}

The branching ratio only depends on the three amplitude combinations:
\begin{equation}\label{eq:Ms}
 M_0 = A_0^{D^0\to V_1V_2}(1+r e^{i\delta}),\qquad
 M_{||} = A^{D^0\to V_1V_2}_{||}(1+ re^{i\delta}),\qquad
 M_\perp = A^{D^0\to V_1V_2}_\perp(1-re^{i\delta}).
\end{equation}
Table~\ref{kpi:weight} shows that $P_{1,2,3}$ yield the relative size and
phase of $M_0$ and $M_{||}$, whereas $P_{4,5,6}$ yield the relative size
and phase of $M_\perp$. Therefore, without further knowledge, one can
extract a combined constraint on $r$, $\delta$ and 
$A_\perp$ from the differential decay width, and one can also
 determine the ratio $A_{||}/A_{0}$. 

Note the invariance under the
simultaneous transformation $A_0A_{||}^*\to A_0^*A_{||}$,
$A_{||}A_\perp^*\to -A_{||}^*A_\perp$, $\delta\to -\delta$, which implies
that for fixed value of $r$ and $A_\perp$ there is a twofold ambiguity on
$\delta$ (in
other words there is no information on the sign of $\sin\delta$).
It is worth noting here that the decay
rate is sensitive to $|\sin\delta|$ terms (thanks to $\Re\left[M_{0,||}
M_\perp^*\right]$), while in the standard
analysis with $PP$ modes one is only sensitive to $\cos\delta$ (neglecting
the small mixing contributions which lift the
ambiguity~\cite{Gronau:2001nr}).
Since $\delta$ is small, the sensitivity on the sine in addition to
the cosine is expected to improve the final result.

The above constraint can be improved by exploiting our current or expected
knowledge of the polarisation of $D\to VV$. If we extract the relative
size and phase of the three amplitudes $A^{D^0\to V_1V_2}_{0,||,\perp}$
from independent single $D\to V_1V_2$ decay (single-tag - ST) measurements,
and if the three amplitudes are not too different in size (as seems to be
the case for $\rho^0 \rho^0$), the measurement of the $M_i$ amplitudes in
the correlated (double-tag - DT) $D\bar D\to (V_1V_2)(K\pi)$ decay leads
to the determination of both $r$ and $\delta$ (more precisely, $r$,
$\cos\delta$ and $|\sin\delta|$). Since the ratio $r$ is already
well known, $r = 0.055\pm 0.002$~\cite{hfwag:2009}, our method may lead
to a good measurement of $\delta$.

Note that for relatively low statistics, a simplified transversity
analysis can be performed. Instead of considering the full angular distribution
in both single and double $D$ decays, one can perform a one-parameter fit
to the distribution of the transversity~\footnote{$(\theta_{V_1},\theta_{\mathrm{tr}},\Phi_\mathrm{tr})$
transversity angles are related to $(\theta_{V_1},\theta_{V_2},\Phi)$
helicity angles by
\begin{equation}
\cos\theta_{V_2}=\sin\theta_\mathrm{tr}\cos\theta_\mathrm{tr}
\qquad
\sin\theta_{V_2}\sin\Phi=\cos\theta_\mathrm{tr} \qquad
\sin\theta_{V_2}\cos\Phi=\sin\theta_\mathrm{tr}\sin\Phi_\mathrm{tr}
\end{equation}
}
angle $\theta_\mathrm{tr}$,
which yields the perpendicular polarisation fraction in
single-tag and double-tag decays:
\begin{equation}
f_\perp^\mathrm{ST}=\frac{|A_\perp|^2}{|A_0|^2+|A_{||}|^2+|A_\perp|^2}\,,\
\ \ 
f_\perp^\mathrm{DT}=\frac{|M_\perp|^2}{|M_0|^2+|M_{||}|^2+|M_\perp|^2}.
\end{equation}
The above observables leads to:
\begin{equation}
\frac{|A_\perp|^2}{|A_0|^2+|A_{||}|^2}=\frac{f_\perp^\mathrm{ST}}{
1-f_\perp^\mathrm{ST}}\,,\ \ \ 
\left|\frac{1+r e^{i\delta}}{1-r
e^{i\delta}}\right|^2=\frac{f_\perp^\mathrm{ST}}{
1-f_\perp^\mathrm{ST}}\frac{1-f_\perp^\mathrm{DT}}{f_\perp^\mathrm{DT}}\,,
\end{equation}
which implies a one-dimensional parabolic constraint in the
$(r,\cos\delta)$ plane. An independent constraint comes from the
ratio of double-tag to single-tag widths proportional to
$(|A_0|^2+|A_{||}|^2+|A_\perp|^2)/(|M_0|^2+|M_{||}|^2+|M_\perp|^2)$,
that can be expressed in terms of $r$, $\cos\delta$ and $f_\perp^\mathrm{ST}$.
This simplified transversity analysis allows one to
determine $r$ and $\cos\delta$. However the main novelty of
our proposal comes from the sensitivity of the complete correlated decay
rate to $|\sin\delta|$ terms, which needs the study of the full angular
dependence.

\subsection{CP-violation in $D^0\bar{D}^0$ mixing}
In the previous discussions, we have neglected the tiny CP-violation in
$D^0\bar{D}^0$ mixing  in order to simplify
the study of correlated $D\to VV$ decays. The inclusion of this effect
would impact our results in the following way: 
\begin{itemize}
\item If CP violation is indeed measured through $\psi(3770)\rightarrow
D^0\bar{D}^0\rightarrow (V_1V_2)(V_3 V_4)$, we cannot \textit{a
 priori} disentangle CP violation in  mixing from CP violation in decay.
Therefore, if we want to convert this result into a bound on fundamental
parameters of a New Physics model, we will have to exploit external inputs
on CP-violating parameters of the mixing (from other observables). On the
other hand, such an input is not necessary if we only aim at setting a
constraint on CP-violation itself.
\item In the determination of the strong phase in $D\to K\pi$, the
amplitudes exhibit in principle a small dependence on mixing effects.
However, this dependence is very weak with respect to the
dependence on the hadronic parameters $(r,\delta)$, and as a first
approximation, it can be neglected. As in the previous case, we can
use external information on the $CP$-violating parameters in mixing to
include their impact when required by more accurate measurements of the
partial decay rate.
\end{itemize}

\section{Potential for BES-III and a super $\tau$-charm factory}
In this section we give a first rough estimate of the expected sensitivity
of
the two different measurements discussed above, either at the BES-III
experiment or at a Super $\tau$-charm factory.

\label{sec:4}
\subsection{$CP$ violation}
\label{sec:4:subsec:2} As discussed in
Section~\ref{sec:3:subsec:2}, the decay chain of $e^+e^- \to \psi
\to D^0 \bar{D}^0 \to f_a f_b$ can be described by
Eq.~(\ref{eq:brcp}), in which both CP conserving and violating
processes can occur. We
parameterize the ratio of amplitudes $\rho_f$ in Eq.~(\ref{rhof}) as
$\rho_f=\eta_f (1+\delta_f) e^{i\alpha_f}$,
where the $\delta_f$ is term from CP violation in decay, and $\alpha_f$
is the phase difference between $D^0$
and $\bar{D}^0$ decay into the same final state $f$.

The $D^0$ decay channels in Table~\ref{tab:1} can be
directly used to search for CP violation by fully considering the
correlation of $D^0\bar{D^0}$ production at BES-III. The background
is small, and the main dilution is due to the mis-identification of
charged particles, which is suppressed by about $10^{-4}$. The
sensitivity of measurement of CP violation can reach about  $10^{-3}$ with
a 20 fb$^{-1}$ luminosity on the $\psi(3770)$ peak at
BES-III.  As in the previous case, one must take care of the background due to
the dilution from non-CP eigenstates that impact
the quasi two-body decays of $D^0$ meson listed in
Table~\ref{tab:2} (for $D \to PV$) and \ref{tab:3} (for $D \to VV$).

Final states consisting of two vector meson pairs are particularly
interesting, since
one can use information on transversity amplitudes to extract different combinations of
CP-violating observables, as
discussed in
Section~\ref{sec:3:subsec:2}: $(0,0)$, $(0,||)$, $ (||,0)$,
$(||,||)$, $(\perp,\perp)$. For example, a back-of-the-envelope
computation yields the most promising channel $\rho^0
\rho^0$/$\bar{K}^{*0} \rho^0$:
\begin{equation}\label{Br alpha}
\mathcal{BR}((D^0\bar{D}^0)_{C=-1} \to \rho^0 \rho^0,\bar{K}^{*0}
\rho^0)\Big|^{CPV}_{(0,||)} \simeq 8 \times
\mathcal{BR}^0(D^0\to\rho^0 \rho^0)\cdot
\mathcal{BR}^{||}(D^0\to\bar{K}^{*0} \rho^0)
 \sin^2\frac{\alpha_a-\alpha_b}{2},
\end{equation}
where $\mathcal{BR}^0$ means the branching fraction for longitudinal
polarized $D^0\to \rho^0\rho^0$ decay, $\mathcal{BR}^{||}$ means
the parallel helicities fraction of $D^0\to \bar{K}^{*0}\rho^0$
decay, and where we have assumed that the CP-violating
parameters $\delta_f$ vanish.

Assuming that no CP-violating signal events in $D^0\bar D^0$
coherent decays are observed with 20 fb$^{-1}$ data at BES-III, we
can provide an upper limit on the CP-violating branching fraction
at 90\% confidence level (C.L.), as indicated in Table~\ref{upper Br}.
A Super $\tau$-charm factory with 2 ab$^{-1}$
data yields naturally stronger constraints. If each polarized
fraction is measured independently,
an upper limit on the phase difference $|\alpha_a-\alpha_b|$ can be set.
For example, the current values for the polarized fractions
in $\rho\rho$ and $\rho K^*$ yields the
upper limit
$|\alpha_a-\alpha_b|<4.4^{\circ}$ at 90\% confidence level from
the channel $(D^0\bar{D}^0)_{C=-1} \to \rho^0 \rho^0,\bar{K}^{*0}
\rho^0\Big|_{(0,||)}$. At a future Super $\tau$-charm factory,
with a data set of 2 ab$^{-1}$, the constraint would be more severe,
$|\alpha_a-\alpha_b|< 0.5^{\circ}$ at 90\% confidence level.

A more realistic analysis requires a likelihood fit to
the full
angular dependence of the $VV$ modes. Systematics will
arise from the mis-reconstruction as $VV$ CP-eigenstates of the events
 that actually come from other resonances or background
contributions. In view of the sizable width of the vector resonances, we
 expect that these systematics will dominate the final result.
Their precise estimate in the framework of each experiment is however
beyond the scope of this paper.

\renewcommand{\multirowsetup}{\centering}
\begin{table}[h]
\begin{center}
\begin{tabular}{cccc}  \hline\hline
\multirow{2}*{Reaction} &\multirow{2}*{Efficiency}  &Upper limits \\&&at BES-III($\times10^{-7}$)\\
 \hline
 $D^0\bar D^0\to (\rho^+\rho^-)(\bar K^{*0}\omega)$    &0.13    &2.46\\
 $D^0\bar D^0\to (\rho^0\rho^0)(\bar K^{*0}\rho^0)$    &0.17    &1.88\\
 $D^0\bar D^0\to (\bar K^{*0}\rho^0)(K^{*0}\omega)$    &0.10    &3.19\\
 $D^0\bar D^0\to (\bar K^{*0}\rho^0)(\rho^0\phi)$      &0.09    &3.55\\
 $D^0\bar D^0\to (\bar K^{*0}\omega)(\rho^0\phi)$      &0.08    &3.99\\
 $D^0\bar D^0\to (\rho^0\rho^0)(\bar K^{*0}\omega)$    &0.15    &2.13\\
 $D^0\bar D^0\to (\rho^0\rho^0)(\rho^0\phi)$           &0.13    &2.46\\
 $D^0\bar D^0\to (\rho^+\rho^-)(\rho^0\phi)$           &0.11    &2.90\\
 $D^0\bar D^0\to (\rho^+\rho^-)(K^{*+}K^{*-})$         &0.11    &2.90\\
 \hline\hline
\end{tabular}
\caption{The projected 90\%-C.L. upper limits on CP violating
branching fraction of some most interesting (VV)(VV) modes from correlated
$D^0\bar D^0$ pairs with 20 fb$^{-1}$ data taken at
$\psi(3770)$ peak at BES-III.}\label{upper Br}
\end{center}
\end{table}

\subsection{Strong phase in $D^0 \rightarrow K\pi$}
\label{sec:4:subsec:1}
The joint decay of $D^0$ into $K^-\pi^+$ and of
$D^0$ into a $CP$ eigenstate $f_{\eta}$  can be described as
\begin{equation}
\Gamma_{K\pi;f_\eta}\equiv \Gamma[(K^- \pi^+)(f_{\eta})] \approx
 A^2A^2_{f_{\eta}}|1+ \eta
r e^{-i \delta} |^2 \approx  A^2A^2_{f_{\eta}}(1+2 \eta r
\cos\delta),
 \label{eq:besiii_delta_rD}
\end{equation}
where $A = |\langle K^- \pi^+ |{\cal H}| D^0 \rangle |$ and
$A_{f_{\eta}} = |\langle f_{\eta} |{\cal H}| D^0 \rangle |$ are the
real-valued decay amplitudes, $\eta = \pm 1$ is CP eigenvalue of the
eigenstate $f_{\eta}$, $r e^{-i \delta}$ is defined in
Eq.(\ref{eq:rdeltaforkpi}) and we have taken $f_{\eta}$ to be a $PP$ or
$VP$ CP-eigenstate, without any non trivial phase-space dependence. We
also have neglected the subdominant $r^2$ term in
Eq.~(\ref{eq:besiii_delta_rD}). The following asymmetry can be used to
determine $\delta$~\cite{liyang:2007}
\begin{eqnarray}
{\cal A} \equiv \frac{\Gamma_{K\pi;f_+} - \Gamma_{K\pi;
f_-}}{\Gamma_{K\pi;f_+} +\Gamma_{K\pi; f_-}},
\label{eq:besiii_delta_rD_a}
\end{eqnarray}
where $\Gamma_{K\pi;f_\pm}$ is defined in
Eq.~(\ref{eq:besiii_delta_rD}), which is the rate for the
$\psi(3770) \rightarrow D^0 \bar{D}^0$ configuration to decay into
flavor eigenstates and a $CP$-eigenstates $f_\pm$.
Eq.~(\ref{eq:besiii_delta_rD}) implies a small asymmetry, ${\cal
A} = 2 r \cos \delta$. In such a case, the error $\Delta {\cal A}$ is
approximately
$1/\sqrt{N_{K^-\pi^+}}$, where $N_{K^-\pi^+}$ is the total number
of events tagged with $CP$-even and $CP$-odd eigenstates, leading to:
\begin{eqnarray}
\Delta (\cos \delta) \approx \frac{1}{2 r
\sqrt{N_{K^-\pi^+}}}. \label{eq:besiii_delta_rD_est}
\end{eqnarray}
The expected number $N_{K^-\pi^+}$ of $CP$-tagged events depends on the
 total number of $D^0 \bar{D}^0$ pairs $N(D^0
\bar{D}^0)$, the branching ratio to the CP-eigenstate $f_\eta$ and the
tagging efficiency. Considering all decay modes listed in
tables~\ref{tab:1}, \ref{tab:2} and \ref{tab:3} we find
\begin{eqnarray}
\Delta (\cos\delta) \approx \frac{ 300}{\sqrt{N(D^0
\bar{D}^0)}}. \label{eq:besiii_delta_rD_est_num}
\end{eqnarray}
At BES-III,  about $72 \times 10^6$ $D^0 \bar{D}^0$ pairs can be
collected with four year running \cite{Asner:2008nq,besiii}, which
implies
an accuracy of about 0.03 for
$\cos\delta$, when considering both $K^- \pi^+$ and $K^+\pi^-$ final
states.

As in the previous section a more realistic analysis requires a likelihood
fit to the full
angular dependence of the $VV$ modes, which in turn provides independent
information on $|\sin\delta|$ as explained above. On the other hand
the imperfect reconstruction of the $VV$ events  as
pure CP-eigenstates will presumably introduce sizable systematics
in this discussion.

At a Super $\tau$-charm factory
\cite{taucharm1,taucharm2} with a 2 ab$^{-1}$ data set, we can expect a
factor of ten improvement, but again the precise impact of the modeling
of the vector resonances requires more studies.

\section{Conclusion}
\label{sec:conclution}

The charm quark offers interesting opportunities to cross-check
 the mechanism of CP violation precisely tested in the strange and beauty
sectors. The start of BES-III will
allow for extensive measurements of charm properties. Among the
various tests that can be considered, one may think of exploiting
the quantum correlations in the $D\bar{D}$ pairs produced at
$\psi(3770)$ resonance. In this paper, we exploit these correlations
in $\psi(3770)\rightarrow D^0\bar{D}^0\rightarrow (V_1V_2)(V_3 V_4)$
in connection with CP violation, and
$\psi(3770)\rightarrow D^0\bar{D}^0\rightarrow (V_1V_2)(K\pi)$ for
CKM angle $\gamma$ measurements, where all $VV$ pairs are reconstructed
as CP-eigenstates.

In the case of $\psi(3770) \rightarrow D^0\bar{D}^0\rightarrow (V_1
V_2)(V_3 V_4)$, the existence of correlations
hinders some helicity configurations for the outgoing vector mesons
in the absence of CP violation. This is mirrored by the angular
distribution of the differential decay width, out of which
CP-violating observables can be constructed. Such observables should
be  interesting to isolate significant New Physics
effects in the charm sector. Assuming that there would be no CP-violating
signal events observed in $D^0\bar D^0$ coherent decays with 20
fb$^{-1}$ data taken at $\psi(3770)$ peak at BES-III, we estimated an
order of magnitude of
the corresponding upper limit on CP-violating parameters, in particular
for the channel $(D^0\bar{D}^0)_{C=-1} \to \rho^0 \rho^0,\bar{K}^{*0} \rho^0)\Big|_{(0,||)}$.
Since the obtained bounds do not follow from a full angular fit and do
not include the systematics corresponding
to the separation of the wanted vector resonances from the background,
further studies are needed.

CP-tagged $D\rightarrow K\pi$ decays give access to
the strong phase difference $\delta$ between Cabibbo-favored and
doubly-Cabibbo suppressed decays, and thus
improve the uncertainty on the $\gamma$ measurement of the unitary
triangle from $B^\pm \rightarrow D/\bar{D} K^\pm$ decays. At BES-III,
with 20 fb$^{-1}$ data at $\psi(3770)$ peak, we estimate the error of
$\cos\delta$ to be of a few percents, corresponding to an error on
the $\delta$ of a few degrees. We expect this estimate can be improved by
taking into account the dependence of the full angular decay width to the
sine of the strong phase. At the
Super $\tau$-charm factory, the expected statistical error on $\delta$
could then fall below one degree. On the other hand a further study of
experimental systematics related to the background identification is
required since they will presumably dominate over the
uncertainty quoted here.

Since our numerical estimates are quite promising we hope that the
potential of such coherent $D$-decays into vector mesons at charm
factories will be assessed more precisely in the future.

\section*{Acknowledgments}

One of the author (H.~B.~Li) would like to thank M.~Z.~Yang and
Z.~Z.~Xing for useful discussions. This work is supported in part by
the ANR contract ANR-06-JCJC-0056, the EU Contract No.
  MRTN-CT-2006-035482, \lq\lq FLAVIAnet'',
the National Natural Science Foundation of China under contracts
Nos. 10521003,10821063,10835001,10979008, the 100 Talents program
of CAS, and the Knowledge Innovation Project of CAS under contract
Nos. U-612 and U-530 (IHEP).

\end{document}